\documentclass[11pt]{article}
\usepackage{fullpage}
\usepackage{color}
\usepackage{citesort}
\usepackage{graphicx}
\def\myendproof{{\ \vbox{\hrule\hbox{%
   \vrule height1.3ex\hskip0.8ex\vrule}\hrule }}\par}
\newtheorem{theorem}{Theorem}
\newtheorem{lemma}{Lemma}

\newenvironment{proof}{{\it Proof. }}{\myendproof}
\newcommand{\floor}[1]{{\left\lfloor{#1}\right\rfloor}}

\newcommand{\setof}[1]{\{{#1}\}}

\newcommand{\Xomit}[1]{}
\newcommand{\leaf}[1]{L_{#1}}
\newcommand{\inter}[1]{{\rm{int}}(#1)}
\newcommand{\score}[2]{{\rm{score}}_{#1}({#2})}
\newcommand{\newscore}[2]{{\rm{score}}_{#1}({#2})}
\newcommand{\hilscore}[1]{{\rm{score}}_{#1}}
\newcommand{\indeg}{\deg^{-}}
\newcommand{\outdeg}{\deg^{+}}

\title{Improved Compact Visibility Representation of Planar Graph via
Schnyder's Realizer\thanks{A preliminary version
is~\cite{LinLS03}. Research supported in part by NSC grant
NSC-91-2213-E-001-028.}}

\author{Ching-Chi Lin \and Hsueh-I Lu\thanks{Corresponding
author. Address: 128 Academia Road, Section 2, Taipei 115, Taiwan.
Email: hil@iis.sinica.edu.tw.
URL: www.iis.sinica.edu.tw/\~{ }hil/.}
\and I-Fan Sun}
\date{Institute of Information Science, Academia Sinica, Taiwan\\\bigskip
\today}
\begin{document}
\maketitle
\begin{abstract}
Let $G$ be an $n$-node planar graph. In a visibility representation of
$G$, each node of $G$ is represented by a horizontal line segment such
that the line segments representing any two adjacent nodes of $G$ are
vertically visible to each other. In the present paper we give the
best known compact visibility representation of $G$.  Given a
canonical ordering of the triangulated $G$, our algorithm draws the
graph incrementally in a greedy manner.  We show that one of three
canonical orderings obtained from Schnyder's realizer for the
triangulated $G$ yields a visibility representation of $G$ no wider
than $\floor{\frac{22n-40}{15}}$.  Our easy-to-implement $O(n)$-time
algorithm bypasses the complicated subroutines for four-connected
components and four-block trees required by the best previously known
algorithm of Kant.  Our result provides a negative answer to Kant's
open question about whether $\floor{\frac{3n-6}{2}}$ is a worst-case
lower bound on the required width.  Also, if $G$ has no degree-three
(respectively, degree-five) internal node, then our visibility
representation for $G$ is no wider than $\floor{\frac{4n-9}{3}}$
(respectively, $\floor{\frac{4n-7}{3}}$).  Moreover, if $G$ is
four-connected, then our visibility representation for $G$ is no wider
than $n-1$, matching the best known result of Kant and He.  As a
by-product, we give a much simpler proof for a corollary of Wagner's
Theorem on realizers, due to Bonichon, Sa\"{e}c, and Mosbah.
\end{abstract}

\section{Introduction}
\label{section:intro}

\begin{figure}
\centerline{\input{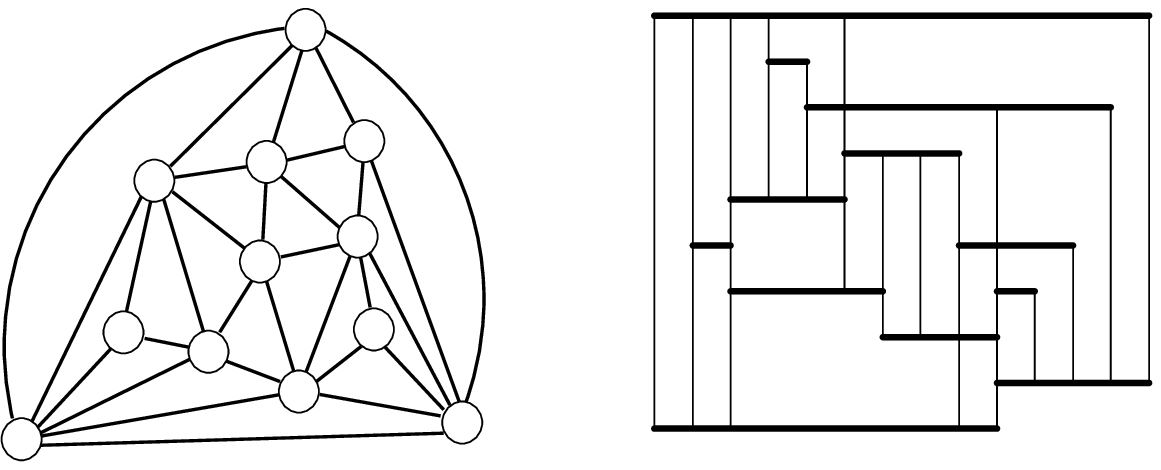}}
\caption{A plane triangulation and one of its visibility
representations.}
\label{fig:visibility}
\end{figure}

In a {\em visibility representation} of a planar graph $G$, the nodes
of $G$ are represented by non-overlapping horizontal line segments, called
{\em node segments}, such that the node segments representing any two
adjacent nodes of $G$ are vertically visible to each other. (See
Figure~\ref{fig:visibility}.)  Computing compact visibility
representations of planar graphs is not only fundamental in
algorithmic graph theory~\cite{TamassiaT89,DiBattistaTT92} but also
practically important in VLSI layout design~\cite{DT85}.

Without loss of generality the input $G$ can be assumed to be an
$n$-node plane triangulation.  Following the convention of placing the
endpoints of node segments on the grid points, one can easily see that
any visibility representation of $G$ can be made no higher than
$n-1$. Otten and van~Wijk~\cite{otten78} gave the first known
algorithm for visibility representations of planar graphs, but no
width bound was provided for the output.  Rosenstiehl and
Tarjan~\cite{tarjan86}, Tamassia and Tollis~\cite{tamassia86}, and
Nummenmaa~\cite{numm92} independently proposed $O(n)$-time algorithms
whose outputs are no wider than $2n-5$.  Kant~\cite{Kant93,Kant97}
improved the required width to at most $\floor{\frac{3n-6}{2}}$ by
decomposing $G$ into its four-connected components and then combining
the visibility representations of the four-connected components into a
visibility representation of $G$.  Kant left open the question of
whether the upper bound $\floor{\frac{3n-6}{2}}$ on the width is also
a worst-case lower bound. In the present paper we provide a negative
answer to Kant's question by presenting an algorithm that always
produces a visibility representation for $G$ whose width is at most
$\floor{\frac{22n-40}{15}}$.

Our algorithm, just like that of Nummenmaa~\cite{numm92}, is based
upon the concept of canonical ordering for plane triangulations.
Specifically, our algorithm draws $G$ incrementally in a greedy manner
according to any given canonical ordering of $G$.  An
arbitrary canonical ordering of $G$ may yield a visibility
representation with width $2n-O(1)$. Rosenstiehl and
Tarjan~\cite{tarjan86} even conjectured that selecting a node ordering
to minimize the area of the corresponding visibility
representation is NP-hard.
We show that the required width can
be bounded by $\floor{\frac{22n-40}{15}}$ using the best one out of the three
canonical orderings obtained from Schnyder's
realizer~\cite{Schnyder90,Schnyder89} for $G$.  
Our algorithm can easily be implemented
to run in $O(n)$ time, bypassing the complicated subroutines of
finding four-connected components and four-block
trees~\cite{KanevskyTDBC91} required by the best previously known
algorithm of Kant~\cite{Kant93,Kant97}.  Also, for the case that $G$
has no degree-three (respectively, degree-five) internal node, the
output visibility representation of our algorithm is no wider than
$\floor{\frac{4n-9}{3}}$ (respectively, $\floor{\frac{4n-7}{3}}$).
Moreover, for the case that $G$ is four-connected, the output
visibility representation of our algorithm is no wider than $n-1$,
matching the best known result due to Kant and
He~\cite{KantH93,KantH97}.

Schnyder's realizer~\cite{Schnyder90,Schnyder89} for plane
triangulation was invented for compact straight-line drawing of plane
graph.
Researchers~\cite{DeFPP90,Kant96,KantH97,DeFraysseixOR94,He99,He01,ChrobakK97,FoessmeierKK97}
also obtained similar and other graph-drawing results using the
concept of canonical ordering for tri-connected plane
graph. Nakano~\cite{Nakano00} attempted to explain the hidden relation
between these two concepts. Recently, Chiang, Lin, and
Lu~\cite{ChiangLL01} presented a new algorithmic tool {\em orderly
spanning tree} that extends the concept of $st$-ordering~\cite{Even76}
(respectively, canonical ordering and realizer) for plane graphs
unrequired to be biconnected (respectively, triconnected and
triangulated).  Orderly spanning tree has been successfully applied to
obtain improved results in compact graph
drawing~\cite{ChiangLL01,LiaoLY01,ChenLLY02}, succinct graph encoding
with query support~\cite{ChiangLL01,ChuangGHKL98}, and design of compact routing
tables~\cite{Lu02}.  Very recently, Bonichon, Gavoille, and
Hanusse~\cite{BonichonGH03} obtained the best known upper bounds on
the numbers of distinct labeled and unlabeled planar graphs based on
{\em well orderly spanning tree}, a special case of orderly spanning
tree.  As a matter of fact, we first successfully obtained the results
of this paper using orderly spanning tree, and then found out
that Schnyder's realizer suffices.

Our analysis requires an equality (see Lemma~\ref{lemma:leaf})
relating the number of internal nodes in the three trees of a realizer
$R$ of $G$ and the number of faces of $G$ intersecting with all three
trees of $R$. The equality was proved very recently by Bonichon,
Sa\"{e}c, and Mosbah~\cite{Bonichon02} as a corollary of the so-called
Wagner's Theorem~\cite{Wagner} on Schnyder's realizers. Their proof
requires a careful case analysis for 32 different configurations. As a
by-product, we give a much simpler proof for the equality without relying
on Wagner's Theorem on realizers.


The remainder of the paper is organized as follows.
Section~\ref{section:preliminaries} gives the preliminaries.
Section~\ref{section:algo} describes and analyzes our algorithm.
Section~\ref{section:lower-bound} discusses the tightness of our
analysis.  Section~\ref{section:conclusion} concludes the paper with
an open question.

\section{Preliminaries}
\label{section:preliminaries}
\begin{figure}
\centerline{\input{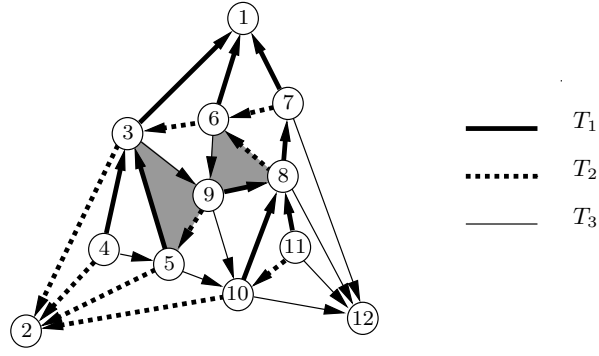}}
\caption{A realizer for the plane triangulation shown in
Figure~\ref{fig:visibility}(a), where $(3,5,9)$ and $(6,9,8)$ are the
only two cyclic faces with respect to this realizer.  The orientation
of each edge is from a child to its parent in the corresponding tree.}
\label{fig:realizer}
\end{figure}

Let $G$ be the input $n$-node {\em plane triangulation}, a planar
graph equipped with a fixed planar embedding such that the boundary of
each face is a triangle. Clearly, $G$ has $2n-5$ internal faces.  Let
$I$ consist of the internal nodes of $G$.  Let $R=(T_1,T_2,T_3)$ be a
{\em realizer} of $G$, which is obtainable in $O(n)$
time~\cite{Schnyder89,Schnyder90}.  That is, the following properties
hold for $R$.
\begin{itemize}
\item
The internal edges of $G$ are partitioned into three edge-disjoint
trees $T_1$, $T_2$, and $T_3$, each rooted at a distinct external node
of $G$.
\item
The neighbors of each node $v$ in $I$ form six blocks $U_1$,
$D_3$, $U_2$, $D_1$, $U_3$, and $D_2$ in counterclockwise order around
$v$, where $U_j$ (respectively, $D_j$) consists of the parent
(respectively, children) of $v$ in $T_j$ for each $j\in\setof{1,2,3}$.
\end{itemize}
For each index $i\in\setof{1,2,3}$, let $\ell_i$ be the node labeling
of $G$ obtained from the counterclockwise preordering of the spanning
tree $\bar{T}_i$ of $G$ consisting of $T_i$ plus the two external
edges of $G$ that are incident to the root of $T_i$. (Each $\bar{T}_i$
is as a matter of fact an orderly spanning tree~\cite{ChiangLL01} of
$G$.)  Let $\ell_i(v)$ be the label of $v$ with respect to $\ell_i$.
For example, Figure~\ref{fig:realizer} shows a realizer of the plane
triangulation shown in Figure~\ref{fig:visibility}(a) with
labeling~$\ell_1$.  The counterclockwise preordering of $\bar{T}_2$ is
$2,12,10,11,5,9,4,3,6,8,7,1$; and that of $\bar{T}_3$ is
$12,1,7,8,11,10,9,6,3,5,4,2$.

\begin{lemma}[see,~e.g.,~\cite{ChiangLL01,numm92,ChuangGHKL98}]
\label{fact:realizer}
The following properties hold for each index $i\in\setof{1,2,3}$,
where $u_1$ and $u_2$ are the nodes with $\ell_i(u_1)=1$ and
$\ell_i(u_2)=2$.
\begin{enumerate}
\item
\label{fact:exterior} The subgraph $G_k$ of $G$ induced by the
nodes $v$ with $1\leq \ell_i(v)\leq k$ is biconnected. The boundary of
$G_k$'s external face is a cycle $C_k$ containing $u_1$ and $u_2$.
\item
\label{fact:interval} If $v$ is the node with $\ell_i(v)=k$, then $v$
is on $C_{k}$; and the neighbors of $v$ in $G_{k-1}$ form an
interval with at least two nodes on the path $C_{k-1}-\setof{(u_1,u_2)}$.
\item
\label{fact:index} The neighbors of $v$ in $G$ form the following four
blocks in counterclockwise order around $v$: $($1$)$ the parent of $v$
in $T_i$, $($2$)$ the node set consists of the neighbors $u$ in
$G-T_i$ with $\ell_i(u)< \ell_i(v)$, $($3$)$ the children of $v$ in
$T_i$, and $($4$)$ the node set consists of the neighbors $u$ in
$G-T_i$ with $\ell_i(u)> \ell_i(v)$.
\end{enumerate}
\end{lemma}
A labeling $\ell$ of $G$ that labels the external nodes by $1$, $2$,
and $n$ and satisfies Lemmas~\ref{fact:realizer}(\ref{fact:exterior})
and~\ref{fact:realizer}(\ref{fact:interval}) is a {\em canonical
ordering} of $G$ (e.g., see~\cite{numm92,Kant96,DeFPP90}).  Therefore,
$\ell_1$, $\ell_2$, and $\ell_3$ are all canonical orderings of $G$.

For each node $v$ of $G$, let $\deg(v)$ denote the degree of $v$, i.e., the number of neighbors
of $v$ in $G$.  For each index $i\in\setof{1,2,3}$, let $\indeg_i(v)$
(respectively, $\outdeg_i(v)$) be the number of neighbors $u$ of $v$
in $G$ with $\ell_i(u)<\ell_i(v)$ (respectively,
$\ell_i(u)>\ell_i(v)$).  Clearly, we have
$\deg(v)=\indeg_i(v)+\outdeg_i(v)$.  For each node $v$ in $I$, let
\begin{eqnarray*}
  \newscore{i}{v}&=&\min\setof{\outdeg_i(v),\indeg_i(v)};\\
  \score{}{v}&=&\newscore{1}{v}+\newscore{2}{v}+\newscore{3}{v}.
\end{eqnarray*}
For example, if $\ell_1$ is the labeling obtained from the tree $T_1$
consisting of the thick edges shown in Figure~\ref{fig:realizer}, then
we have $\newscore{1}{v_8}=2$, $\newscore{1}{v_9}=1$,
$\newscore{1}{v_{10}}=2$, and $\newscore{1}{v_{11}}=1$.
Let
\begin{displaymath}
\hilscore{i}= \sum_{v\in I}\newscore{i}{v}.
\end{displaymath}

Let $[\pi]$ be 1 (respectively, 0) if condition $\pi$ is true
(respectively, false).  Let $\leaf{i}$ consist of the leaves of $T_i$.
For each node $v\in I$, let
\begin{displaymath}
\inter{v}=\sum_{i=1}^{3}[\mbox{$v\not\in L_i$}].
\end{displaymath}
Let $B$ consist of the internal nodes $v$ of $G$ with $\inter{v}=2$
and $\deg(v)=5$.  We have the following lemma.
\begin{lemma}
\label{lemma:inter-bound}
For each node $v$ in $I$, we have $\score{}{v}\geq
3+2\cdot\inter{v}-[v\in B]$.
\end{lemma}
\begin{proof}
By definition of realizer and
Lemma~\ref{fact:realizer}(\ref{fact:index}), it is clear that
\begin{eqnarray*}
\newscore{1}{v}&=&
\min\setof{|D_3|+2,|D_1|+|D_2|+1};\\
\newscore{2}{v}&=&\min\setof{|D_1|+2,|D_2|+|D_3|+1};\\
\newscore{3}{v}&=&\min\setof{|D_2|+2,|D_1|+|D_3|+1}.
\end{eqnarray*}
We may assume without loss of generality that $|D_1|\geq |D_2|\geq
|D_3|\geq 0$.  One can verify the lemma by examining the inequality
for all possible values $0$, $1$, and $2$ of $\inter{v}$.  For
example, if $v\in B$, then we know $\score{}{v}=2+2+2=6$ by
$|D_1|=|D_2|=1$ and $|D_3|=0$. Also, if $\inter{v}=2$ and $v\not\in
B$, then we have $\score{}{v}\geq 2+2+3=7$ by observing $|D_1|\geq 2$,
$|D_2|\geq 1$ and $|D_3|=0$.  The other cases can be verified
similarly.
\end{proof}


An internal face of $G$ is {\em cyclic} if its boundary intersects with
all three trees $T_1$, $T_2$, and $T_3$.  An internal face of $G$ is
{\em acyclic} if it is not cyclic.  For example, in
Figure~\ref{fig:realizer}, faces $(3,5,9)$ and $(6,9,8)$ are
cyclic; all the other internal faces are acyclic.
Let $c$ be the number of cyclic faces of $G$.  The following
lemma was recently proved by Bonichon, Sa\"{e}c, and
Mosbah~\cite{Bonichon02} in an equivalent form. Our alternative proof
is much simpler.

\begin{lemma}[see~\cite{Bonichon02}]
\label{lemma:leaf}
$\sum_{v\in I}\inter{v}=n+c-4$.
\end{lemma}
\begin{proof}
For each index $i\in\setof{1,2,3}$, let ${\rm{int}}_i$ be the number
of internal nodes in $T_i$.  Clearly, $\sum_{v\in
I}\inter{v}=\sum_{i=1}^{3}{\rm{int}}_{i}-3$.  For each node $v\in I$, let $p_{i}(v)$ denote
the parent of $v$ in $T_{i}$.  For each $v\in L_i$, one can verify
that $F_i(v)=(v, p_j(v), p_k(v))$ is an acyclic face of $G$, where
$\setof{i,j,k}=\setof{1,2,3}$. On the other hand, each acyclic face
$(x,y,z)$ of $G$ has to be an $F_i(v)$ for some $v\in\setof{x,y,z}$
and $i\in\setof{1,2,3}$.  By the orientations of the three edges on
$F_i(v)$, one can see that $F_i(v)\ne F_j(u)$ and $F_i(v)\ne F_k(u)$
hold for any node $u\in L_i$.  By $|I|=n-3$, we have
$\sum_{i=1}^{3}|\leaf{i}|=2n-c-5$. Therefore, $\sum_{i=1}^3{\rm{int}}_i
= 3(n-2)-(2n-c-5) = n+c-1$.
\end{proof}

\begin{figure}
\centerline{\input{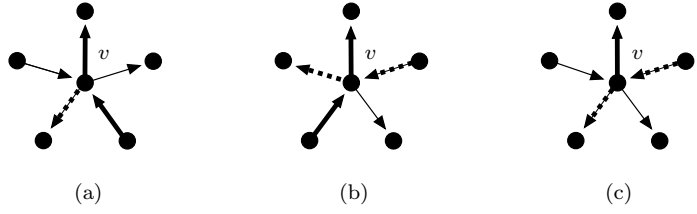}}
\caption{Three different kinds of nodes $v$ in $B$.}
\label{fig:cyclic-face}
\end{figure}

\begin{figure}
\centerline{\input{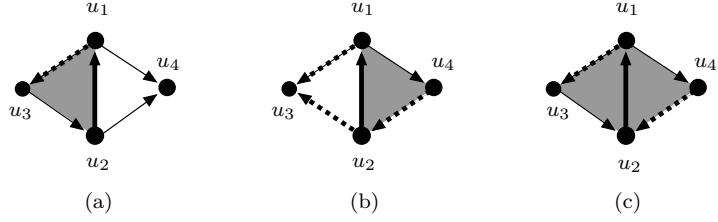}}
\caption{If $u_1$ and $u_2$ are two nodes of $B$ that are adjacent in
$G$, then at least one of faces $(u_1, u_2, u_3)$ and $(u_1, u_2,
u_4)$ is cyclic.}
\label{fig:score-explain}
\end{figure}

\begin{lemma}
\label{lemma:score}\
\begin{enumerate}
\item
\label{item:score3}
If $G$ has no degree-three internal nodes, then
$\sum_{i=1}^{3} \hilscore{i} \geq 5n-15$.
\item
\label{item:score2}
If $G$ has no degree-five internal nodes, then $\sum_{i=1}^{3} \hilscore{i} \geq 5n-17$.
\item
\label{item:score1}
If $G$ is unrestricted, then $\sum_{i=1}^{3} \hilscore{i} \ge \frac{23n}{5}-16$.
\end{enumerate}
\end{lemma}
\begin{proof}
By Lemma~\ref{lemma:inter-bound} we know that if node $v$ in $I$ has
degree more than 3, then $\score{}{v}\geq 5$. By $|I|=n-3$,
Statement~\ref{item:score3} holds.
It follows from Lemmas~\ref{lemma:inter-bound} and~\ref{lemma:leaf}
that $\sum_{i=1}^{3} \hilscore{i} = \sum_{v\in I}
\score{}{v}\geq\sum_{v\in I}3+2 \cdot \inter{v}-[v\in B] =
3(n-3)+2(n+c-4)-|B|$. Therefore, 
\begin{math}
\sum_{i=1}^{3} \hilscore{i} \geq 5n + 2c-|B|-17,
\end{math}
which implies that (a) Statement~\ref{item:score2} holds (by observing
that each node of $B$ has degree five in $G$), and (b)
Statement~\ref{item:score1} can be proved by ensuring $|B|-2c\leq
\frac{2n}{5}-1$ as follows.

Let $k$ be the number of connected components in the subgraph $G[B]$
of $G$ induced by $B$.
Since each of those $2n-5$ internal faces of $G$ is incident to at
most one connected component $G[B]$ and each connected component in
$G[B]$ is incident to at least five faces of $G$, we have $5k\leq
2n-5$.


Let $u_1$ and $u_2$ be two adjacent nodes of $B$ such that $(u_1,u_2)$
is an incoming edge of $u_1$. (That is, $u_1$ is the parent of $u_2$
in some tree $T_i$ of $R$.) Let $(u_3,u_1,u_2)$ and $(u_4,u_1,u_2)$ be
the two faces of $G$ that contain edge $(u_1,u_2)$.  One can see that
at least one of faces $(u_1,u_2,u_3)$ and $(u_1,u_2,u_4)$ is cyclic by
verifying, with the assistance of Figure~\ref{fig:cyclic-face}, that
(a) both edges $(u_1,u_3)$ and $(u_1,u_4)$ have to be outgoing from
$u_1$; and (b) at least one of edges $(u_3,u_2)$ and $(u_4,u_2)$ is
incoming to $u_2$, as illustrated by Figure~\ref{fig:score-explain}.
Let $F$ be an arbitrary spanning forest of $G[B]$, which clearly has
$|B|-k$ edges. Each cyclic face contains at most two edges of $F$ and
each edge of $F$ is incident to at least one cyclic face. Thus, we
have $|B|-k \le 2c$.
\end{proof}

\section{Our algorithm}
\label{section:algo}
\begin{figure}
\centerline{\input{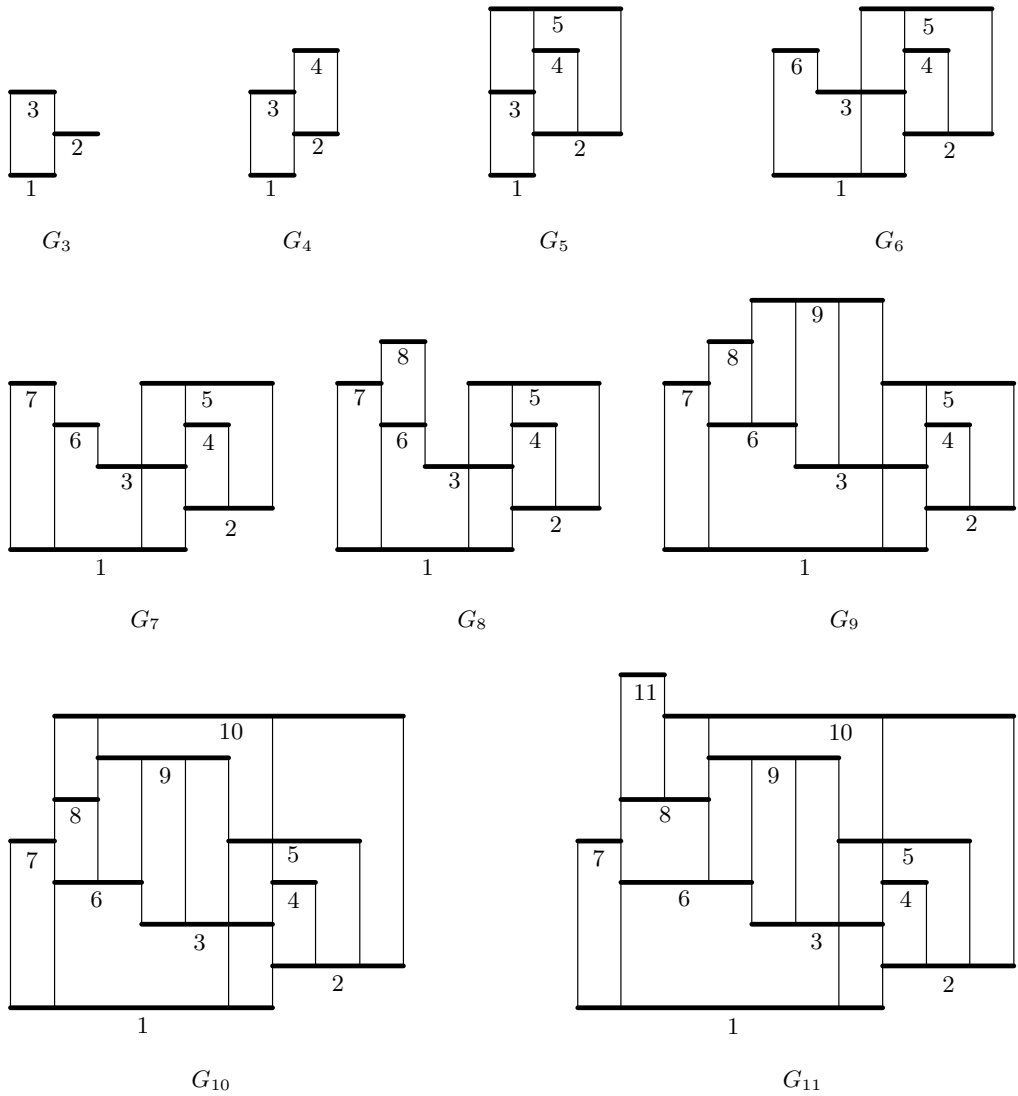}} 
\caption{The intermediate steps of our algorithm for obtaining the
visibility representation for the plane triangulation shown in
Figure~\ref{fig:visibility}(a) with respect to the canonical ordering
specified by its node labels.}
\label{fig:drawing}
\end{figure}

Let $\ell_i$ be a given canonical ordering of the input $n$-node plane
triangulation $G$.  For each $k=1,2,\ldots,n$, let $v_k$ be the node
with $\ell_i(v_k)=k$ and let $G_k$ be the subgraph of $G$ induced by
$v_1,v_2,\dots,v_k$.  Clearly, $G_3$ is a triangle and $v_1$, $v_2$,
and $v_n$ are the external nodes of $G$.  Our algorithm initially
produces a visibility representation of $G_3$ as shown in
Figure~\ref{fig:drawing}, and then extends that into a visibility
representation of $G=G_n$ in $n-3$ iterations: For each
$k=4,5,\ldots,n$, the $(k-3)$-rd iteration obtains a visibility
representation of $G_k$ from that of $G_{k-1}$ by 
\begin{enumerate}
\item extending the
visibility representation of $G_{k-1}$ in a greedy manner until the
node segment of each neighbor of $v_k$ is visible from above, and then
\item placing the shortest possible node segment representing $v_k$
from above that yields a visibility representation of $G_k$.
\end{enumerate}
For example, if $G$ is as shown in Figure~\ref{fig:visibility}(a) and
$\ell_i$ is as specified by the node labels, then the visibility
representations for $G_3,G_4,\ldots,G_{11}$ are as shown in
Figure~\ref{fig:drawing} and the resulting visibility representation
of $G=G_{12}$ is as shown in Figure~\ref{fig:visibility}(b). The
correctness of our algorithm follows from the fact that $\ell_i$ is a
canonical ordering of $G$, which therefore satisfies
Lemma~\ref{fact:realizer}(\ref{fact:exterior}).  A naive
implementation of our algorithm takes $O(n^2)$ time. However, it is
not difficult to implement our algorithm to run in $O(n)$ time using
basic data structures like doubly linked lists to support $O(1)$-time
operations such as determining whether a node segment is visible from
above and inserting a new column of grid points.



%
%
%

\begin{theorem}\label{theorem:main}
Any $n$-node plane triangulation $G$ with $n>3$ has an $O(n)$-time
obtainable visibility representation whose width is at most
\begin{enumerate}
\item
\label{thm:item2}
 $\floor{\frac{4n-9}{3}}$, if $G$ has no degree-three internal nodes;
\item
\label{thm:item1}
$\floor{\frac{4n-7}{3}}$, if $G$ has no degree-five internal nodes; or
\item
\label{thm:item3} $\floor{\frac{22n-40}{15}}$, if $G$ is
unrestricted.
\end{enumerate}
\end{theorem}
\begin{proof}
By Lemma~\ref{lemma:score} and the fact that a realizer is obtainable
in linear time, it suffices to show that the width of the output visibility
representation by our algorithm is at most $3n-8 -\sum_{v\in
I}\newscore{i}{v}$. For each $k=4,5,\ldots,n$, consider
the iteration that produces the visibility
representation of $G_k$.  Let $v_j$ be any neighbor of $v_k$ in $G_k$.
In the first half of the iteration, if the node segment of $v_j$ does
not contain any grid point that is visible from above, then a new
column of grid points is inserted to ensure that the node segment for
$v_j$ is visible from above; otherwise, the number of grid points on
the node segment of $v_j$ that are visible from above stays the same.
%
In the second half of the iteration, if $v_k$ is the neighbor of $v_j$
with the largest index, then the node segment of $v_j$ can no longer be
visible from above for the remaining iterations of our algorithm;
otherwise, the number of grid points on the node segment of $v_j$ that
are visible from above decreases by exactly one. Moreover, the node
segment of $v_k$ contains at least $\indeg_i(v_k)$ grid points that
are visible from above in the resulting visibility representation of
$G_k$.
Therefore, in the first half of those $\outdeg_i(v_k)$ iterations, one
for each neighbor $v_\ell$ of $v_k$ with $\ell>k$, 
at most $\outdeg_i(v_k)-\newscore{i}{v_k}$ new columns of grid points
are inserted. Note that $n>3$ implies $\newscore{i}{v_3}\geq 1$.  It
follows that the resulting visibility representation for $G_n=G$ has
width at most
$2+\left(-3+\sum_{k=1}^{n-1}\outdeg_i(v_k)\right)-\left(2+\sum_{k=4}^{n-1}\newscore{i}{v_k}\right)\geq
3n-8-\sum_{v\in I}\newscore{i}{v}$.
\end{proof}

The following result was first obtained by Kant and
He~\cite{KantH93,KantH97} based upon their linear-time algorithm for
obtaining a canonical ordering $\ell_i$ for any $n$-node four-connected
plane triangulation such that $\deg_{i}^{+}(v)\geq 2$ and
$\deg_{i}^{-}(v)\geq 2$ hold for $n-4$ out of the $n-3$ internal nodes
$v$ of $G$. We can alternatively prove the theorem in a much simpler
way: According to the proof of Theorem~\ref{theorem:main}, the width
of the output visibility representation by our algorithm is at most
$3n-8-\sum_{v\in I}\newscore{i}{v} \leq n-1$.

\begin{theorem}[see~\cite{KantH93,KantH97}]
\label{corollary-1}
If $G$ is an $n$-node four-connected plane triangulation, then there is
an $O(n)$-time obtainable visibility representation for $G$ whose
width is at most $n-1$.
\end{theorem}

%

\begin{figure}
\centerline{\input{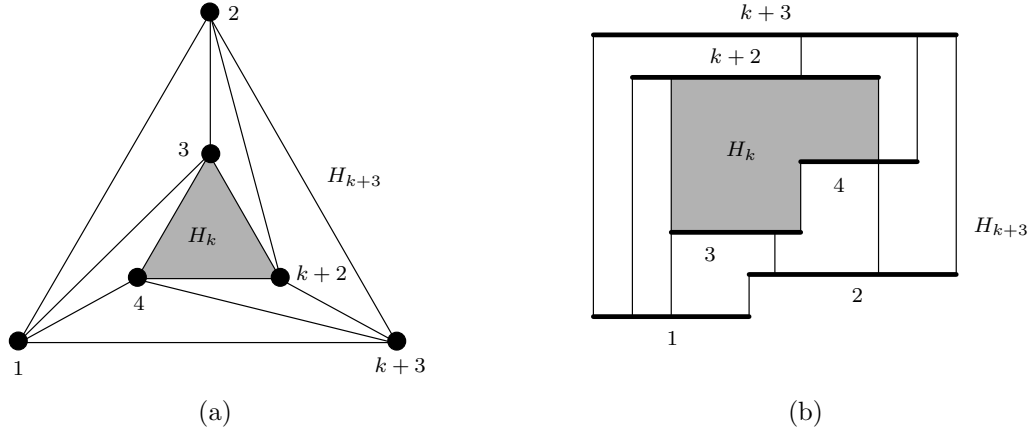}}
\caption{An example showing that our analysis on the required width is
almost tight.}
\label{fig:g6}
\end{figure}

\section{Near tightness of our analysis}
\label{section:lower-bound}

The following lemma shows that our analysis on the required width is
almost tight.
\begin{lemma}
\label{theorem:lower-bound}
For any $n\geq 3$, there exists an $n$-node plane triangulation $H_n$
such that any visibility representations of $H_n$ obtained by our
algorithm with respect to any canonical ordering of $H_n$ has width at
least $\floor{\frac{4n-8}{3}}$.
\end{lemma}
\begin{proof}
We prove the lemma by induction on $n$.  Let $H_3$ (respectively,
$H_4$ and $H_5$) be a plane triangulation with $3$ (respectively, $4$
and $5$) nodes.  Clearly, any visibility representation of $H_3$
(respectively, $H_4$ and $H_5$) has width at least $2$ (respectively,
$3$ and $4$), so the lemma holds for $n=3,4,5$.  For each index $k\geq
3$, let $H_{k+3}$ be the $(k+3)$-node plane triangulation obtained
from $H_k$ by adding three new external nodes and triangulating the
faces as shown in Figure~\ref{fig:g6}(a).  By
Lemma~\ref{fact:realizer}(\ref{fact:exterior}), if $\ell$ is a
canonical ordering of $H_{k+3}$, then the ordering $\ell'$ with
$\ell'(v)=\ell(v)-2$ for each node $v$ of $H_k$ remains a canonical
ordering of $H_{k}$.  As illustrated in Figure~\ref{fig:g6}(b), it is
not difficult to see that the visibility representation for $H_{k+3}$
produced by our algorithm with respect to any canonical ordering of
$H_{k+3}$ is at least 4 units wider than that of $H_{k}$ produced by
our algorithm with respect to any canonical ordering of
$H_k$. Therefore, the lemma is proved.
\end{proof}



\section{Concluding remarks}
\label{section:conclusion}
Whether our upper bound $\frac{22n}{15}-\Theta(1)$ on the required
width is worst-case optimal remains open.


\bibliographystyle{abbrv}
\bibliography{visibility}
%
%
%
%
\end{document}